\documentclass[aps,pre,twocolumn,showpacs,superscriptaddress]{revtex4}
\usepackage{graphicx}

\begin{document}

\title{Early appraisal of the fixation probability in directed networks}

\author{Valmir C. Barbosa}
\affiliation{Programa de Engenharia de Sistemas e Computa\c c\~ao, COPPE,
Universidade Federal do Rio de Janeiro,
Caixa Postal 68511, 21941-972 Rio de Janeiro - RJ, Brazil}

\author{Raul Donangelo}
\affiliation{Instituto de F\'\i sica,
Universidade Federal do Rio de Janeiro,
Caixa Postal 68528, 21941-972 Rio de Janeiro - RJ, Brazil}
\affiliation{Instituto de F\'\i sica, Facultad de Ingenier\'\i a,
Universidad de la Rep\'ublica,
Julio Herrera y Reissig 565, 11.300 Montevideo, Uruguay}

\author{Sergio R. Souza}
\affiliation{Instituto de F\'\i sica,
Universidade Federal do Rio de Janeiro,
Caixa Postal 68528, 21941-972 Rio de Janeiro - RJ, Brazil}
\affiliation{Instituto de F\'\i sica,
Universidade Federal do Rio Grande do Sul,
Caixa Postal 15051, 91501-970 Porto Alegre - RS, Brazil}

\begin{abstract}
In evolutionary dynamics, the probability that a mutation spreads through the
whole population, having arisen in a single individual, is known as the fixation
probability. In general, it is not possible to find the fixation probability
analytically given the mutant's fitness and the topological constraints that
govern the spread of the mutation, so one resorts to simulations instead.
Depending on the topology in use, a great number of evolutionary steps may be
needed in each of the simulation events, particularly in those that end with the
population containing mutants only. We introduce two techniques to accelerate
the determination of the fixation probability. The first one skips all
evolutionary steps in which the number of mutants does not change and thereby
reduces the number of steps per simulation event considerably. This technique is
computationally advantageous for some of the so-called layered networks. The
second technique, which is not restricted to layered networks, consists of
aborting any simulation event in which the number of mutants has grown beyond a
certain threshold value, and counting that event as having led to a total
spread of the mutation. For large populations, and regardless of the network's
topology, we demonstrate, both analytically and by means of simulations, that
using a threshold of about $100$ mutants leads to an estimate of the fixation
probability that deviates in no significant way from that obtained from the
full-fledged simulations. We have observed speedups of two orders of magnitude
for layered networks with $10\,000$ nodes.
\end{abstract}


\maketitle

\section{Introduction}
\label{sec:intro}

We consider directed networks in which each node is inhabited by a single
individual of a population and whose edges represent the possibilities for an
individual's offspring to replace some other individual. Such networks provide
the substrate on which the evolution of the population can be studied given the
constraints imposed by their structures. In this modality of evolutionary
dynamics, known as evolutionary graph dynamics since its introduction in
\cite{lhn05}, the population evolves in discrete time steps, each of which
involves the fitness-based selection of an individual for reproduction and the
use of its offspring to replace one of its out-neighbors in the network. The
chief quantity one targets in such studies is the probability that a mutation
arising at a randomly chosen individual, henceforth called a mutant, of the
otherwise homogeneous population eventually spreads through all the population.
This probability is known as the fixation probability.

In the last decade, the study of several other phenomena has been approached
from a similar perspective of interacting agents. Such phenomena have included
differently constrained forms of the dynamics of evolution \cite{m58,vm07}, the
spread of epidemics through populations \cite{bbpv04}, the emergence of
cooperation in biological and social systems \cite{sp05,ohln06,tdw07,fwnh09},
and various others \cite{gh05,sf07,pc09}. In most cases, what the interacting
agents do, driven by either competition or the goal of promoting cooperation, is
to spread information through the network in order to attempt to influence the
states of other agents. In general, network structure is a major player in
affecting the global outcome of such interactions, and this holds to the extent
that subtle structural changes can have relevant consequences
\cite{kthmcc02,lhn05,gwhxw06,wxw06,onp07,opn07,ss07}. The importance of network
structure, in fact, is also central in several other areas, as for example those
discussed in \cite{bs03,blmch06,nbw06}.

\begin{figure}[t]
\includegraphics[scale=0.30]{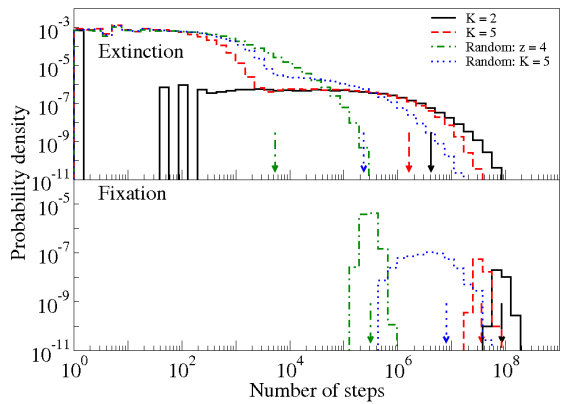}
\caption{(Color online) Probability densities associated with the number of
steps required for extinction and fixation of a mutant $10\%$ fitter than the
remainder of the population. Data are log-binned to the base $1.5$ and represent
averages over $10^6$ events for the $K$-funnel, $10^7$ events for the random
networks ($10^4$ events for each of $10^3$ graphs; in the case of the random
network with a Poisson-distributed number of out-neighbors, each of these
graphs has at least $95\%$ of the nodes in the GSCC). Arrows indicate the means.
All the networks have $1\,555$ nodes.
}
\label{fig:nsteps}
\end{figure}

The fixation probability is very heavily influenced by the structure of the
underlying network as well. In rare cases it is possible to calculate it
analytically from both structure and the relative fitnesses of the individuals
\cite{lhn05,ars06,br08,sar08,mo09}, but in general one has to resort to
simulations of the evolutionary steps. Given the network and the mutant's
fitness, the simulation is conducted as a number of independent events, each of
which starts by placing the mutant at a randomly chosen node and then carries
out the evolutionary steps until either fixation (all nodes contain mutants) or
extinction (no node contains a mutant) occurs. The fraction of events ending in
fixation is an estimate of the fixation probability. This simulation-based
approach to obtaining the fixation probability can be very time-consuming, not
only because many independent events are needed, but also because each event can
require a significantly large number of steps to converge. This is illustrated
in Fig.~\ref{fig:nsteps}, which suggests two properties of the simulation
process. The first is that events ending in extinction usually require
substantially fewer steps to conclude than those ending in fixation. The second
is that the number of steps required for fixation varies widely with network
topology.

The various network topologies used in Fig.~\ref{fig:nsteps} recur throughout
the paper, so we pause momentarily to introduce them. They are the $K$-funnel
\cite{lhn05}, a random generalization thereof \cite{bds09}, and the directed
variant of the Erd\H{o}s-R\'{e}nyi random graphs \cite{er59} discussed in
\cite{k90,bds03}. For $b,K>1$ integers, the $K$-funnel of base $b$ has node set
partitioned into $K$ subsets, called layers and numbered $0,1,\ldots,K-1$, such
that there are $b^k$ nodes in layer $k$. The $K$-funnel, therefore, has
$(b^K-1)/(b-1)$ nodes. For $k=1,2,\ldots,K-1$, an edge exists directed from each
node in layer $k$ to each node in layer $k-1$. Additionally, an edge exists
directed from the single node in layer $0$ to each of the nodes in layer $K-1$.
The $K$-funnel is generalized by any network in which nodes occupy layers and
edges exist only from all nodes in a layer to all nodes in the next according to
some cyclic arrangement of the layers. The generalization we use \cite{bds09} is
constructed randomly by first placing one node at each layer and then adding one
extra node at a time to a layer that is chosen randomly with a bias that favors
those layers that are nearest to the layer having the most nodes ``upstream''
from them. The other random networks we use \cite{k90,bds03} are such that a
randomly chosen node has a number of out-neighbors given by the Poisson
distribution of mean $z>1$ [so a giant strongly connected component (GSCC) is
expected to exist with high probability]. The simulation of the evolutionary
dynamics on such a network is confined to its GSCC.

We also remark, before proceeding, that some of our own recent work related to
the fixation probability has been strongly influenced by the computational
difficulties associated with estimating it. For example, in \cite{bds09} we set
out to create a randomized growing mechanism for layered networks that would
result in fitness amplifiers in the sense of \cite{lhn05}. The mechanism we
created gives rise to the $K$-layer random networks introduced above and,
indeed, we were able to demonstrate the desired amplification effect for
$1\,000$ nodes. However, we fell short of demonstrating to full completion that
the same holds for significantly larger networks, due mainly to the very large
times required to compute the fixation probabilities of the thousands of
candidate networks.

Our central concern in this paper is the devising of simulation strategies that
can make the calculation of the fixation probability substantially faster while
maintaining accuracy or reducing it only imperceptibly. We proceed along two
tracks. The first is targeted at eliminating the evolutionary steps at which no
change is effected to the number of mutants in the network. This occurs whenever
the simulation prescribes that a mutant's offspring is to replace another mutant
or that a non-mutant's offspring is to replace another non-mutant. Our results
in this track are better suited to the layered networks introduced above; we
demonstrate them for the $K$-funnel. The second track we pursue builds on the
realization that, if in general it takes a lot more steps for fixation to occur
than for extinction, then detecting early in the course of a simulation event
that fixation is highly likely to occur can be used as a surrogate to the
eventual detection of fixation and thereby reduce the number of necessary steps.
We have found that, nearly regardless of the network for a large number of
nodes, there exists a threshold number of mutants beyond which fixation is
practically guaranteed to take place. We give results for a wide variety of
networks.

We organize the remainder of the paper in the following manner. In
Section~\ref{sec:fp} we briefly review the key notions related to the fixation
probability. Then in Sections~\ref{sec:layered} and \ref{sec:allnets} we pursue
the two tracks outlined above for computing the fixation probability more
efficiently, respectively for layered networks and for unrestricted networks.
We conclude in Section~\ref{sec:concl}.

\section{Fixation probability}
\label{sec:fp}

Let $P$ be a population of $n$ individuals and, for $i\in P$, let
$P_i\subseteq P\setminus\{i\}$ be the set of individuals that an offspring of
$i$ can replace during the evolution of $P$. Let also $Q_i$ be the set of
individuals whose offspring can replace $i$. These give rise to a directed
graph, called $D$, whose set of nodes is $P$ and whose set of edges, denoted by
$E$, contains the edge $(i,j)$ if and only if $j\in P_i$ (equivalently,
$i\in Q_j$). Each individual $i$ has a fitness $f_i>0$ associated with it, and
similarly to each edge $(i,j)$ there corresponds a probability $w_{ij}$ such
that $\sum_{j\in P_i}w_{ij}=1$ for all $i\in P$. The dynamics of evolution that
we consider occurs in a sequence of steps. At each step, an individual $i$ is
chosen with probability proportional to $f_i$, then another individual
$j\in P_i$ is chosen with probability $w_{ij}$, and finally $j$ is replaced by
an offspring of $i$ having fitness $f_i$.

The fixation probability of $D$, denoted by $\rho$, is the probability that a
mutation spreads through all of $P$ given that it arises at one single
individual and that all individuals in the remainder of the population have the
same fitness. The value of $\rho$ depends on the structure of $D$ and on the
ratio $r$ of the mutant's fitness to that of the other individuals. Moreover, it
is the relationship between $\rho$ and $r$ that determines whether evolution is
driven primarily by natural selection or by random drift: essentially, natural
selection predominates when $\rho$ and $r$ are highly correlated, random drift
otherwise. Note in this context that, if $D$ is not strongly connected (i.e.,
there exist nodes $i$ and $j$ such that no directed path leads from $i$ to $j$),
then $\rho>0$ if and only if there exists an individual from which all others
are reachable. This may cause random drift to be the main driver of evolution,
so henceforth we assume that $D$ is strongly connected (thus $\rho>0$
necessarily).

This type of evolutionary dynamics can be described by a discrete-time Markov
chain of states $0,1,\ldots,n$, each representing a possible number of mutants
in $D$. In this chain, states $0$ and $n$ are absorbing and all others are
transient. If $s$ is a transient state, then from $s$ it is possible to move to
state $s+1$ (with probability $p_s$), to state $s-1$ (with probability $q_s$),
or to remain at state $s$ (with probability $1-p_s-q_s$). If we denote by
$P_{n+1}(n\mid s)$ the probability that, having started at state $s$, this
$(n+1)$-state system eventually enters state $n$, then it is well-known that
\begin{equation}
P_{n+1}(n\mid s)=\frac
{1+\sum_{u=1}^{s-1}\prod_{v=1}^{u}q_v/p_v}
{1+\sum_{u=1}^{n-1}\prod_{v=1}^{u}q_v/p_v}
\label{eq:abs-prob}
\end{equation}
(cf.\ \cite{e00} and references therein), whence $\rho$, which is given by
$P_{n+1}(n\mid 1)$, is such that
\begin{equation}
\rho=\frac
{1}
{1+\sum_{u=1}^{n-1}\prod_{v=1}^{u}q_v/p_v}.
\label{eq:fix-prob}
\end{equation}

When the probabilities $w_{ij}$ are such that $\sum_{j\in Q_i}w_{ji}=1$ for
all $i\in P$ (i.e., not only do the probabilities associated with the outgoing
edges of $i$ sum up to $1$, but also those of the incoming edges), and only
then, the isothermal theorem of \cite{lhn05,n06} establishes that $p_v/q_v=r$
for all $v\in\{1,2,\ldots,n-1\}$. In this case, and assuming that $r\neq 1$
(i.e., the mutation is either advantageous or disadvantageous, but not neutral),
it follows from Eq.~(\ref{eq:fix-prob}) that the fixation probability, now
denoted by $\rho_1$, is
\begin{equation}
\rho_1=\frac
{1-1/r}
{1-1/r^n}.
\label{eq:moran}
\end{equation}
This includes the Moran process \cite{m58}, in which $D$ is such that
$P_i=P\setminus\{i\}$ for all $i\in P$ and $w_{ij}=1/\vert P_i\vert$ for all
$(i,j)\in E$ (we use $\vert X\vert$ to denote the cardinality of set $X$). It
also includes the more general case in which the $w_{ij}$'s are thus constrained
and, moreover, all nodes have as many incoming edges as outgoing edges, provided
this number is the same for all nodes. Other noteworthy cases that also employ
these locally uniform $w_{ij}$'s are the directed graphs that in \cite{lhn05}
are shown to be fitness amplifiers with respect to the expression for $\rho_1$.
One example is the $K$-funnel, for which substituting $r^K$ for $r$ in
Eq.~(\ref{eq:moran}) yields $\rho=\rho_K$ as $n\to\infty$, where
\begin{equation}
\rho_K=\frac
{1-1/r^K}
{1-1/r^{Kn}}.
\label{eq:amplified}
\end{equation}

\section{Fixation in layered networks}
\label{sec:layered}

We henceforth use $w_{ij}=1/\vert P_i\vert$ for all $(i,j)\in E$ exclusively.
For $t\ge 0$ an integer, let $E(t)\subset E$ be the set of edges $(i,j)$ such
that, at the $t$th evolutionary step, node $i$ is a mutant but node $j$ is not.
If $P^+(t)$ denotes the probability that, at step $t+1$, the number of mutants
in $D$ increases (necessarily by $1$), then
\begin{equation}
P^+(t)=\sum_{(i,j)\in E(t)}\frac
{r/\vert P_i\vert}
{n+M(t)(r-1)},
\end{equation}
where $M(t)$ is the number of mutants at step $t$ [so $M(0)=1$]. Letting
$m_i(t)$ be an indicator of whether node $i$ is a mutant at step $t$ [i.e.,
$m_i(t)=1$ in the affirmative case, $0$ otherwise], this expression can be
rewritten as a sum over all edges in $E$:
\begin{eqnarray}
P^+(t)
&=&
\frac{r}{n+M(t)(r-1)}
\sum_{(i,j)\in E}\frac{m_i(t)[1-m_j(t)]}{\vert P_i\vert}\cr
&=&
\frac{r}{n+M(t)(r-1)}\cr
&&\left[
\sum_{(i,j)\in E}\frac{m_i(t)}{\vert P_i\vert}
-\sum_{(i,j)\in E}\frac{m_i(t)m_j(t)}{\vert P_i\vert}
\right].
\end{eqnarray}
And since $D$ is strongly connected by assumption, we obtain
\begin{equation}
P^+(t)=
\frac{r}{n+M(t)(r-1)}
\left[
M(t)
-\sum_{(i,j)\in E}\frac{m_i(t)m_j(t)}{\vert P_i\vert}
\right].
\label{eq:P+}
\end{equation}
Similarly, the probability $P^-(t)$ that the number of mutants decreases
(necessarily by $1$) at step $t+1$ is
\begin{eqnarray}
P^-(t)
&=&
\frac{1}{n+M(t)(r-1)}\cr
&&\left[
\sum_{(i,j)\in E}\frac{m_j(t)}{\vert P_i\vert}
-\sum_{(i,j)\in E}\frac{m_i(t)m_j(t)}{\vert P_i\vert}
\right].
\label{eq:P-}
\end{eqnarray}
The number of mutants remains unchanged at step $t+1$ with probability
$1-P^+(t)-P^-(t)$.

These expressions for $P^+(t)$ and $P^-(t)$ allow the evolutionary dynamics to
be observed from some interesting perspectives. For example, any $D$ for which
$\rho=\rho_1$ is such that
$\sum_{(i,j)\in E}m_j(t)/\vert P_i\vert=
\sum_{j\in P}m_j(t)\sum_{i\in Q_j}1/\vert P_i\vert=
\sum_{j\in P}m_j(t)=M(t)$,
and therefore $P^+(t)/P^-(t)=r=p_s/q_s$ for any state $s\in\{1,2,\ldots,n-1\}$
such that $M(t)=s$. This, essentially, is the argument behind the isothermal
theorem. Moreover, for $r$ sufficiently close to $1$, what makes $P^+(t)$ and
$P^-(t)$ differ from each other is the balance between $M(t)$ and
$\sum_{(i,j)\in E}m_j(t)/\vert P_i\vert$, where the former does not depend on
the topology of $D$ (given $t$) while the latter does. Thus, for example, if we
consider the $K$-funnel and let $M_k(t)$ be the number of mutants in layer $k$
at step $t$, we have
\begin{eqnarray}
\sum_{(i,j)\in E}\frac{m_j(t)}{\vert P_i\vert}
&=&\sum_{k=0}^{K-2}\frac{M_k(t)b^{k+1}}{b^k}+\frac{M_{K-1}(t)}{b^{K-1}}\cr
&=&b[M(t)-M_{K-1}(t)]+\frac{M_{K-1}(t)}{b^{K-1}}.
\end{eqnarray}
Readily, maintaining a relatively high value for the ratio $P^+(t)/P^-(t)$ (a
strong forward bias) in the case of the $K$-funnel depends crucially on how
close $M(t)$ and $M_{K-1}(t)$ are to each other, i.e., on how close the largest
layer is to containing a significant fraction of the $M(t)$ mutants. The fact
that $P^+(t)/P^-(t)\to r^K$ as $n\to\infty$ indicates that the topology of the
$K$-funnel is in fact successful at maintaining the necessary distribution of
mutants.

Equations~(\ref{eq:P+}) and (\ref{eq:P-}) are also useful in that they provide
an alternative mechanism for simulating the evolutionary dynamics. Instead of
repeatedly choosing $i$, then $j\in P_i$ to receive $i$'s offspring, until the
mutation either spreads through the whole of $D$ or dies out, we use the two
equations to decide, at each step $t$, whether the number of mutants will
increase, decrease, or remain the same at step $t+1$. In the former two cases
we choose the nodes to be involved, create or destroy a mutant as the case may
be, then compute $P^+(t+1)$ and $P^-(t+1)$. By doing so, all steps in which no
mutant is created or destroyed are skipped. Of course, in order for this
alternative to be computationally attractive the suppression of these steps has
to compensate for the additional effort to calculate the probabilities at every
step that is not suppressed.

While it is unclear that this will be so in the general case, for layered
networks like the $K$-funnel, in which all nodes in the same layer are
topologically identical to one another (they all have the same in- and
out-neighbors), the simulation can be conducted by keeping track only of the
number of mutants in each layer and the alternative becomes attractive. We then
consider the generalization of the $K$-funnel obtained by letting layer $k$ have
any number $n_k>0$ of nodes. In this case, we can rewrite Eqs.~(\ref{eq:P+}) and
(\ref{eq:P-}) by decomposing each of $P^+(t)$ and $P^-(t)$ into $K$ summands,
each related to a pair of subsequent layers. Thus, we obtain
$P^+(t)=\sum_{k=0}^{K-1}P_k^+(t)$ and $P^-(t)=\sum_{k=0}^{K-1}P_k^-(t)$, with
\begin{equation}
P_k^+(t)=\frac{rM_{k+1}(t)[n_k-M_k(t)]}{n_k[n+M(t)(r-1)]}
\label{eq:Pk+}
\end{equation}
and
\begin{equation}
P_k^-(t)=\frac{M_k(t)[n_{k+1}-M_{k+1}(t)]}{n_k[n+M(t)(r-1)]},
\label{eq:Pk-}
\end{equation}
where $M_k(t)$ is as for the $K$-funnel. In these expressions, and henceforth,
layer numbers are incremented or decremented modulo $K$.

For layered networks such as these, once it has been decided that a mutant is to
be created or destroyed at step $t+1$, the layer at which this is to happen is
$k$ with probability proportional to $P_k^+(t)$, in the case of creation, or
$P_k^-(t)$, in the case of destruction [consistently with this, notice that
$P_k^+(t)=0$ if $M_{k+1}(t)=0$ or $M_k(t)=n_k$, and that $P_k^-(t)=0$ if
$M_k(t)=0$ or $M_{k+1}(t)=n_{k+1}$]. And once $M_k(t+1)$ has been updated from
$M_k(t)$, it follows from Eqs.~(\ref{eq:Pk+}) and (\ref{eq:Pk-}) that only
$P_k^+(t+1)$ and $P_{k-1}^+(t+1)$, or $P_k^-(t+1)$ and $P_{k-1}^-(t+1)$, need to
be calculated. This is so because, although $M(t+1)$ is updated from $M(t)$ as
well, the only effect this has is to alter the factor in the denominator of
Eqs.~(\ref{eq:Pk+}) and (\ref{eq:Pk-}) that is common to all layers. As a
consequence, the probabilities corresponding to any layer other than $k$ or
$k-1$ need not be calculated.

\begin{figure}[t]
\includegraphics[scale=0.30]{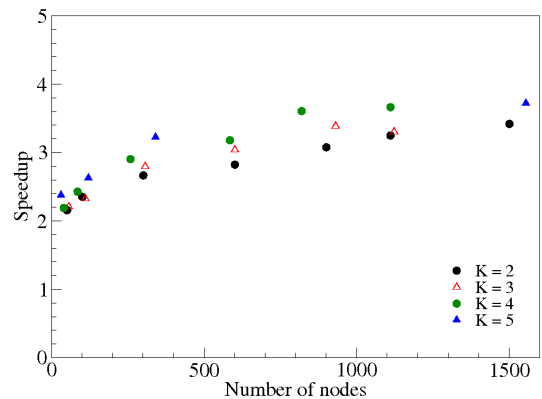}
\caption{(Color online) Speedups resulting from the use of Eqs.~(\ref{eq:Pk+})
and (\ref{eq:Pk-}) on the $K$-funnel with $r=1.1$. Data are averages over at
least $10^4$ events.}
\label{fig:speedupK1}
\end{figure}

Computational results on the $K$-funnel are given in Fig.~\ref{fig:speedupK1}
for a variety of $n$ and $K$ values, where we show the speedup afforded by the
use of Eqs.~(\ref{eq:Pk+}) and (\ref{eq:Pk-}) to compute the fixation
probability. This speedup is defined in terms of processor time and indicates,
in all cases examined, a reduction to less than half the time required by the
simulation that goes through all the evolutionary steps. We note, however, that
the approach we discuss in Section~\ref{sec:allnets} allows for much more
significant speedups.

\section{Fixation in arbitrary networks}
\label{sec:allnets}

In the absence of the computational facilitation provided by layered networks,
which as we have seen allows the fixation probability to be computed more
efficiently by skipping all steps of the simulation in which no mutant is
created or destroyed, for an unrestricted topology we turn to the alternative
strategy of attempting an early stop of each simulation event based on how many
mutants there are. The central question is whether there exists a threshold
number of mutants which, once crossed from below, ensures that fixation is bound
to occur with probability as close to $1$ as one wishes. We provide an
affirmative answer in what follows.

The probability that the mutation eventually dies out, given that $s$ mutants
are originally present, equals $1-P_{n+1}(n\mid s)$. We denote it by
$Q_{n+1}(0\mid s)$, and it follows from Eq.~(\ref{eq:abs-prob}) that
\begin{equation}
Q_{n+1}(0\mid s)=\frac
{\sum_{u=s}^{n-1}\prod_{v=1}^{u}q_v/p_v}
{1+\sum_{u=1}^{n-1}\prod_{v=1}^{u}q_v/p_v}.
\label{eq:ext-prob}
\end{equation}
We are interested in the probability that, conditioned on the fact that
extinction does actually occur, the number of mutants eventually grows from the
initial $s$ to some fixed value $M\in\{s,s+1,\ldots,n-1\}$ but does not surpass
it. We denote this probability by $Q_{n+1}^M(0\mid s)$ and remark that, should
its dependency with $M$ be known, we would immediately be able to discover the
desired threshold for the number of mutants by adopting $s=1$ and specifying a
lower bound on the probability. In other words, we would discover the threshold,
call it $M^*$, by specifying $Q^*$ such that $Q_{n+1}^M(0\mid 1)\ge Q^*$ for all
$M\le M^*$. This follows from the intuitive expectation that
$Q_{n+1}^M(0\mid s)$ is to decrease as $M$ grows for sufficiently large $M$.

From its definition as a conditional probability, $Q_{n+1}^M(0\mid s)$ is given
by $AB/Q_{n+1}(0\mid s)$, where $A$ is the probability that the number of
mutants in the network eventually increases from $s$ to $M$ and $B$ is the
probability that, given that it has $M$ mutants, the system eventually returns
to state $0$ without ever increasing its number of mutants beyond $M$. We
calculate the values of $A$ and $B$ by resorting to discrete-time Markov chains
entirely analogous to the one we have been using, but now having reduced numbers
of states. The first of these chains has states $0,1,\ldots,M$, of which $0$ and
$M$ are absorbing while all else remains unchanged. We set $A$ to the
probability that the system gets absorbed into state $M$ having started at state
$s$, that is, $A=P_{M+1}(M\mid s)$. The second chain has states
$0,1,\ldots,M+1$, with $0$ and $M+1$ the absorbing states and everything else
unchanged. We set $B$ to the probability that absorption occurs at state $0$
once the system is started at state $M$, that is, $B=Q_{M+2}(0\mid M)$. We then
obtain
\begin{equation}
Q_{n+1}^M(0\mid s)=\frac
{P_{M+1}(M\mid s)Q_{M+2}(0\mid M)}
{Q_{n+1}(0\mid s)}.
\label{eq:M-ext-prob}
\end{equation}

\begin{figure}[t]
\includegraphics[scale=0.30]{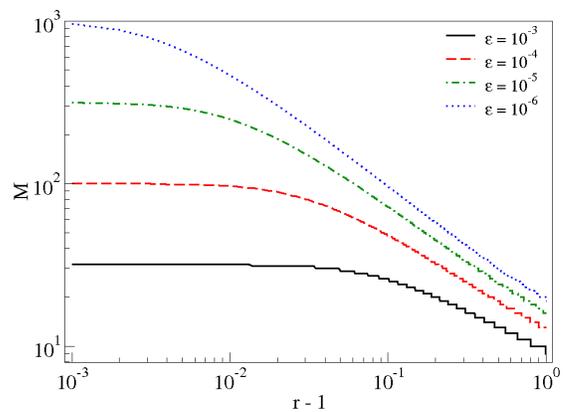}
\caption{(Color online) $M$ as a function of $r$ for
$Q_{n+1}^M(0\mid 1)=\epsilon$.}
\label{fig:M(r)}
\end{figure}

Closed-form expressions are in general not known for Eqs.~(\ref{eq:abs-prob}) or
(\ref{eq:ext-prob}), so we assume that it suffices to consider the case in which
$p_v/q_v=r$ for all transient states $v$. In this case, and for $r\neq 1$,
Eq.~(\ref{eq:M-ext-prob}) yields
\begin{equation}
Q_{n+1}^M(0\mid s)=
\frac{(1-1/r^s)(1-1/r)}{(1-1/r^M)(1-1/r^{M+1})r^M}
\frac{1-1/r^n}{1/r-1/r^n}.
\label{eq:M-ext-prob-c}
\end{equation}
Readily, $Q_{n+1}^M(0\mid s)$ becomes independent of $n$ as $n$ grows,
regardless of whether $r>1$ or $r<1$ (but note that the two limits differ). For
$n\to\infty$, in Fig.~\ref{fig:M(r)} we show plots of
$Q_{n+1}^M(0\mid 1)=\epsilon$ for different values of $\epsilon$ and
$1<r\le 2$. As expected, for fixed $\epsilon$ the value of $M$ increases with
decreasing $r$ in this range, and somewhat counterintuitively we see that the
rate of increase is ever smaller as $r$ approaches $1$. However, this is easily
confirmed as we realize that, as in the figure, the limit $Q$ of
$Q_{n+1}^M(0\mid 1)$ as $r\to 1$ is such that $M\approx\sqrt{1/Q}$.

\begin{figure}[t]
\includegraphics[scale=0.30]{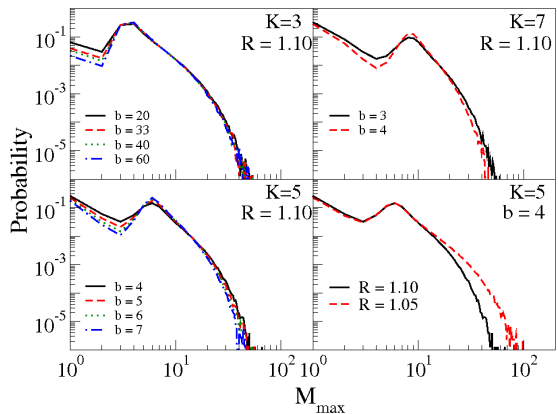}
\caption{(Color online) Probability distributions of $M$ for the $K$-funnel.
Data are averages over $10^6$ events.}
\label{fig:MmaxK}
\end{figure}

\begin{figure}[t]
\includegraphics[scale=0.30]{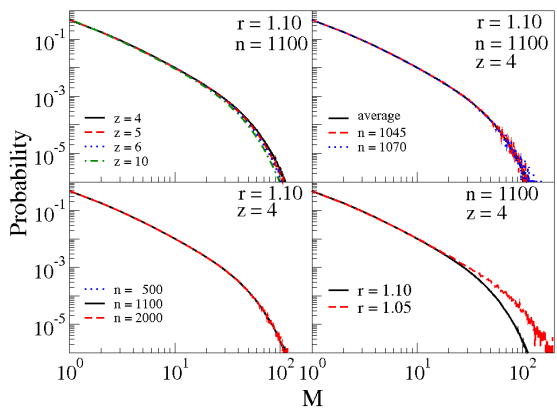}
\caption{(Color online) Probability distributions of $M$ for random networks
with a Poisson-distributed number of out-neighbors. Data are averages over
$10^7$ events ($10^4$ events for each of $10^3$ graphs with at least $0.95n$
nodes in the GSCC). The plot labeled ``average'' refers to all data for $r=1.1$,
$n=1\,100$, and $z=4$ in all panels. This is compared in the same panel to the
results corresponding to a single graph whose GSCC has $1\,045$ or $1\,070$
nodes, for which $10^6$ events have been run.}
\label{fig:MmaxER}
\end{figure}

\begin{figure}[t]
\includegraphics[scale=0.30]{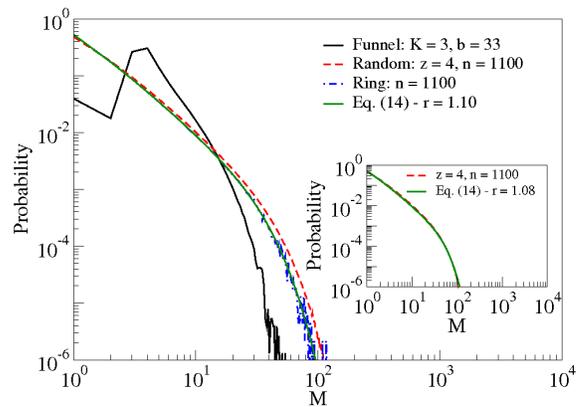}
\caption{(Color online) Probability distributions of $M$ for assorted
topologies with $r=1.1$. Data are averages over $10^6$ events for the $K$-funnel
and the unidirectional ring, $10^7$ events for the random network ($10^4$ events
for each of $10^3$ graphs with at least $0.95n$ nodes in the GSCC).}
\label{fig:Mmax}
\end{figure}

Computational results on the distribution of $M$, the maximum number of mutants
achieved when the dynamics ends in extinction, are given in
Figs.~\ref{fig:MmaxK} through \ref{fig:Mmax} for a variety of topologies.
Figure~\ref{fig:MmaxK} refers to the $K$-funnel, Fig.~\ref{fig:MmaxER} to random
networks with a Poisson-distributed number of out-neighbors, and
Fig.~\ref{fig:Mmax} to a selection of networks that includes an instance of each
of these two types and also the unidirectional ring. The latter figure also
includes plots of the analytical prediction given by
Eq.~(\ref{eq:M-ext-prob-c}). In this respect, notice that the prediction that
corresponds to the $r$ value used in the simulations matches the data for the
unidirectional ring perfectly. This, of course, is consistent with the fact
that, in this case, the requirements of the isothermal theorem are satisfied and
therefore the assumption behind Eq.~(\ref{eq:M-ext-prob-c}) is exact. Similarly,
in the figure's inset we demonstrate that, for the random network in use, the
prediction of Eq.~(\ref{eq:M-ext-prob-c}) with a slightly lower value of $r$ is
also a perfect match.

\begin{figure}[t]
\includegraphics[scale=0.30]{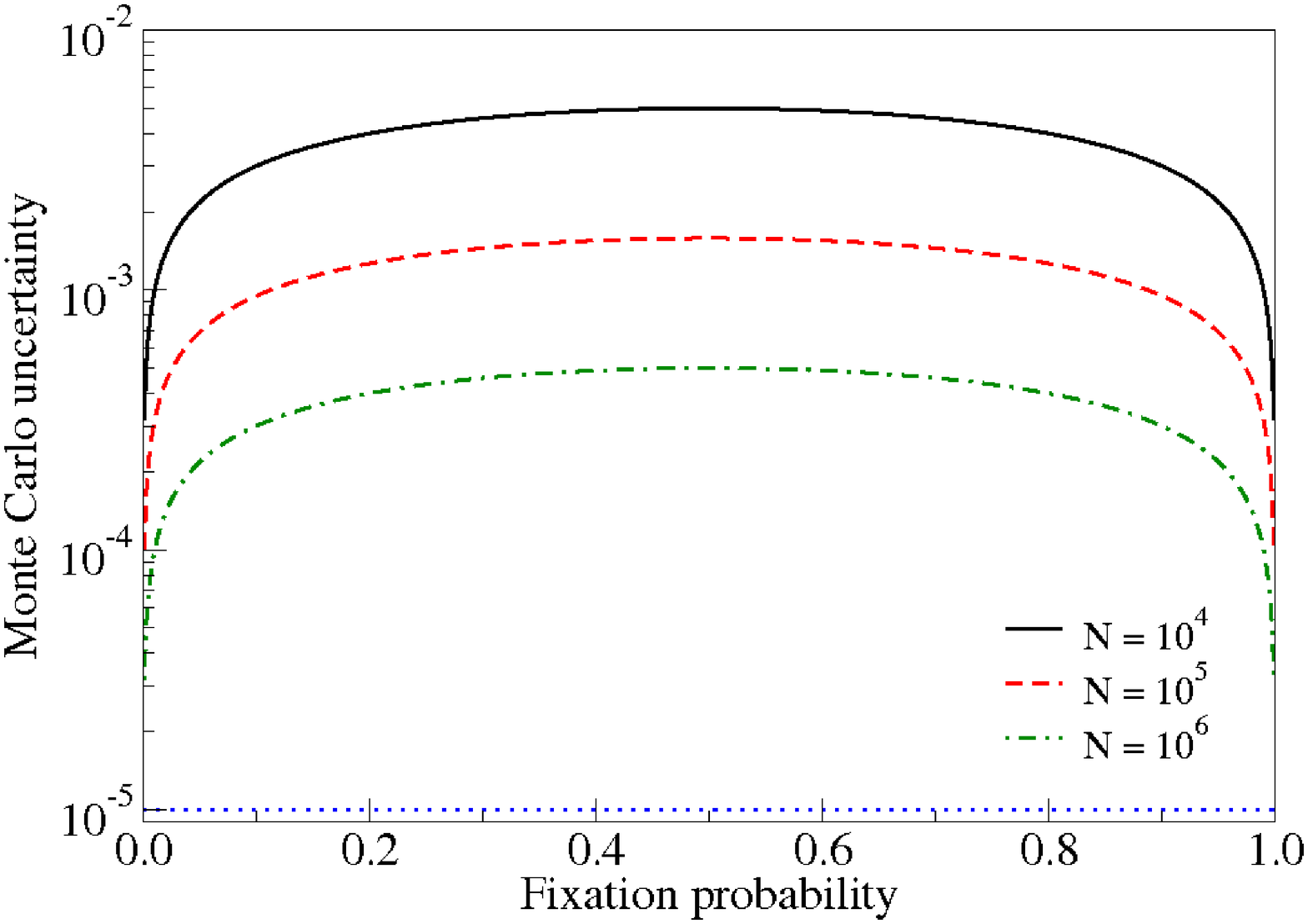}
\caption{(Color online) Inherent uncertainty associated with the Monte Carlo
computations of the fixation probability, as given by the standard error of the
fixation probability estimate $\hat\rho$.}
\label{fig:error}
\end{figure}

Notice also, in all three figures, that for $M=100$ no probability is above
$10^{-5}$. This means that, during the simulations, of all events that ended in
extinction, no more than one out of $10^5$ achieved more than $100$ mutants. As
a consequence, should we use $M^*=100$ (corresponding roughly to $Q^*=10^{-6}$,
given that there were at least $10^6$ events) and count as an event ending in
fixation any event achieving more than $M^*$ mutants, we would be introducing a
deviation of no more than $10^{-5}$ with respect to the actual value of the
fixation probability. But the fixation probability obtained by full simulations
of the evolutionary dynamics is itself subject to the so-called standard error
that is inherent to any Monte Carlo simulation. If $\hat\rho$ denotes the
fixation probability calculated after $N$ events, then the standard error is the
standard deviation of the $0$'s (extinctions) and $1$'s (fixations) accumulated
along the events divided by $\sqrt{N}$, that is,
$\sqrt{\hat\rho(1-\hat\rho)/N}$. This function of $\hat\rho$ is plotted in
Fig.~\ref{fig:error} for different values of $N$, along with a flat line for the
constant $10^{-5}$. Clearly, the additional deviation introduced by the use of
$M^*=100$ is negligible when compared to the standard error.

\begin{figure}[t]
\includegraphics[scale=0.30]{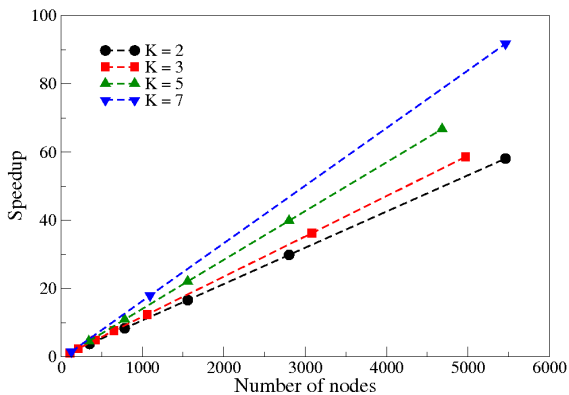}
\caption{(Color online) Speedups resulting from the use of the $M^*=100$
threshold on the $K$-funnel with $r=1.1$. Data are averages over $10^6$ events.}
\label{fig:speedupK2}
\end{figure}

\begin{figure}[t]
\includegraphics[scale=0.30]{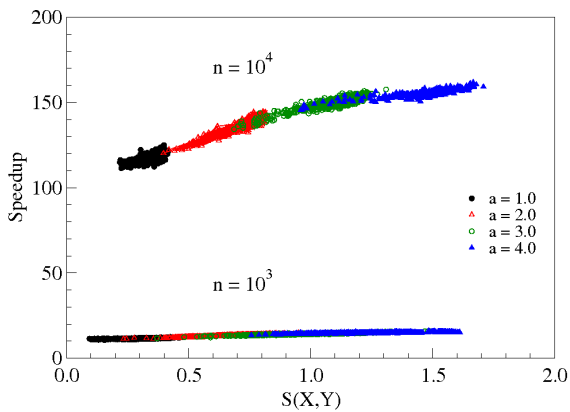}
\caption{(Color online) Speedups resulting from the use of the $M^*=100$
threshold on the random networks of \cite{bds09} with $r=1.1$. Data are averages
over $10^7$ events ($10^4$ events for each of $10^3$ graphs), except for
$n=10\,000$ with $a=4$, in which case $10^4$ events are used for each of $500$
graphs.}
\label{fig:speedupOur}
\end{figure}

Speedup figures resulting from the use of early fixation detection for $M^*=100$
are shown in Figs.~\ref{fig:speedupK2} and \ref{fig:speedupOur}, respectively
for the $K$-funnel (with several $n$ and $K$ values) and the $K$-layer random
networks of \cite{bds09} (with two values of $n$, $K=5$, and several values of
the $a$ parameter that in \cite{bds09} is used to control the layer-selection
mechanism as the network is grown by the addition of new nodes). Plots in the
latter figure are given against $S(X,Y)$, which in \cite{bds09} is used to
indicate, if sufficiently above $0$, how close each network is to having, like
the $K$-funnel, an exponentially growing number of nodes per layer as one moves
``upstream'' through the layers. Clearly, speedups are very significant,
particularly for the $K$-funnel with the largest values of $n$ and $K$ and all
random networks of $10\,000$ nodes. We also remark that, in
Fig.~\ref{fig:speedupK2}, the fact that slopes increase with $K$ is closely
related to Eq.~(\ref{eq:amplified}): as the fraction of events that lead to
fixation grows with $K$, we expect from Fig.~\ref{fig:nsteps} that the average
number of steps behave likewise; qualitatively, this explains why speedups
increase with $K$.

\section{Final remarks}
\label{sec:concl}

\begin{figure}[t]
\includegraphics[scale=0.30]{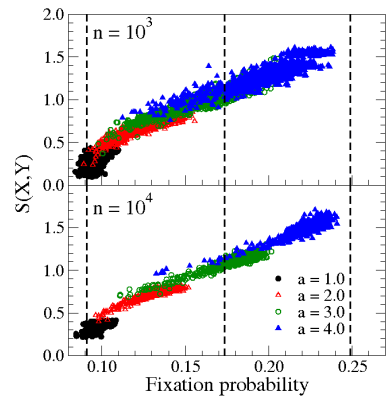}
\caption{(Color online) Results for the random networks of \cite{bds09} with
$K=5$ and $r=1.1$. Each graph whose layer populations correlate with those of
the $K$-funnel by more than $0.9$ in the sense of the Pearson correlation
coefficient is represented by its fixation probability and its $S(X,Y)$ value.
Data are averages over $10^7$ events ($10^4$ events for each of $10^3$ graphs),
except for $n=10\,000$ with $a=4$, in which case $10^4$ events are used for each
of $500$ graphs. Dashed lines indicate $\rho=\rho_1$ through $\rho_3$.}
\label{fig:fpOur}
\end{figure}

In Section~\ref{sec:intro} we mentioned that, in \cite{bds09}, we were unable
to extend to larger values of $n$ our conclusions regarding the
fitness-amplification properties of the $K$-layer random networks we used in
some of this paper's experiments. Provisioned with the technique of
Section~\ref{sec:allnets}, we can now bypass the computational difficulties that
hampered our progress in that occasion by employing early detection of fixation.
Doing this for $M^*=100$ has resulted in the data shown in Fig.~\ref{fig:fpOur},
from which it is finally clear that, also for $n=10\,000$, it is possible to
grow layered networks that achieve significant fitness amplification. As we see
in the figure, for $K=5$ many grown networks have $\rho$ values between $\rho_2$
and $\rho_3$. We note, however, that this is still an easier scenario than that
of \cite{bds09}, where $K=10$ was used for $n=10\,000$, since in the present
case we had to calculate the speedups given in Fig.~\ref{fig:speedupOur} and
these required that simulation events be carried out to completion.

Another interesting by-product of our use of the threshold number of mutants
$M^*$ is that, should the dynamics be started with $s>M^*$ randomly placed
mutants of equal fitness, then fixation would occur almost surely. This is so
because the probability of there being so many mutants in a dynamics that is
bound to extinction is as small as allowed by the choice of $Q^*$. However, note
that Eq.~(\ref{eq:M-ext-prob-c}) is of no immediate help in quantifying the
``almost surely,'' since it is conditioned on the dynamics ending in extinction
and therefore does not apply to those cases in which the number of mutants
becomes large enough that extinction is unlikely. Nevertheless, whenever the
isothermal theorem holds, Eq.~(\ref{eq:abs-prob}) implies that fixation from the
initial $s$ mutants occurs with probability $(1-1/r^s)/(1-1/r^n)$ (see also
\cite{lhn05}), which is asymptotically equal to $1$ for $r>1$ as both $s$ and
$n$ grow.

It is also worth mentioning that, because this study has been targeted at
directed graphs, Eqs.~(\ref{eq:P+}) and (\ref{eq:P-}) are also applicable to the
special case of an undirected graph and can lead to useful insight also in this
case. Specifically, suppose we take any strongly connected $D$ without
antiparallel edges and make it functionally undirected by adding to it the
antiparallel counterpart of every one of its edges. If $E$ and the $P_i$ sets
continue to refer to the original $D$, then the contribution of each mutant $l$
to $\sum_{(i,j)\in E}m_j(t)/\vert P_i\vert$ in Eq.~(\ref{eq:P-}) jumps from
$\sum_{(i,l)\in E}1/\vert P_i\vert$ to
$\sum_{(i,l)\in E}1/\vert P_i\vert+\sum_{(l,i)\in E}1/\vert P_i\vert$, therefore
leading to a smaller $P^+(t)/P^-(t)$ ratio. On the other hand, if $D$ already
has antiparallel edges, then one curious special case is that of the
$2$-superstar of \cite{lhn05}, which has a central node with $n-1$ peripheral
nodes that connect to it through antiparallel edge pairs. This graph is already
functionally undirected and, for $n\to\infty$, $\rho\to\rho_2$. So the
$2$-superstar is somewhat of an exception with regard to the $P^+(t)/P^-(t)$
ratio for undirected graphs. In fact, thus far we have been unable to find any
other undirected graphs for which $\rho\ge\rho_2$, but to the best of our
knowledge the question of whether any exist remains unsettled.

We note, finally, that fixation for the $2$-superstar, when it happens, is bound
to take a considerable number of steps to occur (this can be seen in
Fig.~\ref{fig:nsteps}, since the $2$-superstar and the $2$-funnel with the same
number of nodes are the same graph). For the $2$-superstar, the sum from
Eq.~(\ref{eq:P-}) mentioned above is
$M_\mathrm{p}(t)/(n-1)+m_\mathrm{c}(t)(n-1)$, where $M_\mathrm{p}(t)$ is the
number of peripheral mutants at step $t$ and $m_\mathrm{c}(t)$ is either $1$ or
$0$, indicating respectively whether or not a central mutant exists at that
step. Clearly, for $r$ only slightly above $1$, obtaining $P^+(t)>P^-(t)$
depends crucially on the existence of the central mutant. This, in turn, can
easily change from step to step until fixation is eventually approached. It is
precisely in cases such as this that the early estimates of the fixation
probability introduced in Section~\ref{sec:allnets} are most useful.

\begin{acknowledgments}
We acknowledge partial support from CNPq, CAPES, a FAPERJ BBP grant, the joint
PRONEX initiative of CNPq and FAPERJ under contract 26.171.528.2006, and
CNPq-PROSUL.
\end{acknowledgments}

\bibliography{fastfp}

\begin{thebibliography}{30}
\expandafter\ifx\csname natexlab\endcsname\relax\def\natexlab#1{#1}\fi
\expandafter\ifx\csname bibnamefont\endcsname\relax
  \def\bibnamefont#1{#1}\fi
\expandafter\ifx\csname bibfnamefont\endcsname\relax
  \def\bibfnamefont#1{#1}\fi
\expandafter\ifx\csname citenamefont\endcsname\relax
  \def\citenamefont#1{#1}\fi
\expandafter\ifx\csname url\endcsname\relax
  \def\url#1{\texttt{#1}}\fi
\expandafter\ifx\csname urlprefix\endcsname\relax\def\urlprefix{URL }\fi
\providecommand{\bibinfo}[2]{#2}
\providecommand{\eprint}[2][]{\url{#2}}

\bibitem[{\citenamefont{Lieberman et~al.}(2005)\citenamefont{Lieberman, Hauert,
  and Nowak}}]{lhn05}
\bibinfo{author}{\bibfnamefont{E.}~\bibnamefont{Lieberman}},
  \bibinfo{author}{\bibfnamefont{C.}~\bibnamefont{Hauert}}, \bibnamefont{and}
  \bibinfo{author}{\bibfnamefont{M.~A.} \bibnamefont{Nowak}},
  \bibinfo{journal}{Nature} \textbf{\bibinfo{volume}{433}},
  \bibinfo{pages}{312} (\bibinfo{year}{2005}).

\bibitem[{\citenamefont{Moran}(1958)}]{m58}
\bibinfo{author}{\bibfnamefont{P.~A.~P.} \bibnamefont{Moran}},
  \bibinfo{journal}{Proc. Camb. Phil. Soc.} \textbf{\bibinfo{volume}{54}},
  \bibinfo{pages}{60} (\bibinfo{year}{1958}).

\bibitem[{\citenamefont{Valleriani and Meene}(2007)}]{vm07}
\bibinfo{author}{\bibfnamefont{A.}~\bibnamefont{Valleriani}} \bibnamefont{and}
  \bibinfo{author}{\bibfnamefont{T.}~\bibnamefont{Meene}},
  \bibinfo{journal}{Ecol. Model.} \textbf{\bibinfo{volume}{208}},
  \bibinfo{pages}{159} (\bibinfo{year}{2007}).

\bibitem[{\citenamefont{Barth\'{e}lemy
  et~al.}(2004)\citenamefont{Barth\'{e}lemy, Barrat, Pastor-Satorras, and
  Vespignani}}]{bbpv04}
\bibinfo{author}{\bibfnamefont{M.}~\bibnamefont{Barth\'{e}lemy}},
  \bibinfo{author}{\bibfnamefont{A.}~\bibnamefont{Barrat}},
  \bibinfo{author}{\bibfnamefont{R.}~\bibnamefont{Pastor-Satorras}},
  \bibnamefont{and}
  \bibinfo{author}{\bibfnamefont{A.}~\bibnamefont{Vespignani}},
  \bibinfo{journal}{Phys. Rev. Lett.} \textbf{\bibinfo{volume}{92}},
  \bibinfo{pages}{178701} (\bibinfo{year}{2004}).

\bibitem[{\citenamefont{Ohtsuki et~al.}(2006)\citenamefont{Ohtsuki, Hauert,
  Lieberman, and Nowak}}]{ohln06}
\bibinfo{author}{\bibfnamefont{H.}~\bibnamefont{Ohtsuki}},
  \bibinfo{author}{\bibfnamefont{C.}~\bibnamefont{Hauert}},
  \bibinfo{author}{\bibfnamefont{E.}~\bibnamefont{Lieberman}},
  \bibnamefont{and} \bibinfo{author}{\bibfnamefont{M.~A.} \bibnamefont{Nowak}},
  \bibinfo{journal}{Nature} \textbf{\bibinfo{volume}{441}},
  \bibinfo{pages}{502} (\bibinfo{year}{2006}).

\bibitem[{\citenamefont{Santos and Pacheco}(2005)}]{sp05}
\bibinfo{author}{\bibfnamefont{F.~C.} \bibnamefont{Santos}} \bibnamefont{and}
  \bibinfo{author}{\bibfnamefont{J.~M.} \bibnamefont{Pacheco}},
  \bibinfo{journal}{Phys. Rev. Lett.} \textbf{\bibinfo{volume}{95}},
  \bibinfo{pages}{098104} (\bibinfo{year}{2005}).

\bibitem[{\citenamefont{Taylor et~al.}(2007)\citenamefont{Taylor, Day, and
  Wild}}]{tdw07}
\bibinfo{author}{\bibfnamefont{P.~D.} \bibnamefont{Taylor}},
  \bibinfo{author}{\bibfnamefont{T.}~\bibnamefont{Day}}, \bibnamefont{and}
  \bibinfo{author}{\bibfnamefont{G.}~\bibnamefont{Wild}},
  \bibinfo{journal}{Nature} \textbf{\bibinfo{volume}{447}},
  \bibinfo{pages}{469} (\bibinfo{year}{2007}).

\bibitem[{\citenamefont{Fu et~al.}(2009)\citenamefont{Fu, Wang, Nowak, and
  Hauert}}]{fwnh09}
\bibinfo{author}{\bibfnamefont{F.}~\bibnamefont{Fu}},
  \bibinfo{author}{\bibfnamefont{L.}~\bibnamefont{Wang}},
  \bibinfo{author}{\bibfnamefont{M.~A.} \bibnamefont{Nowak}}, \bibnamefont{and}
  \bibinfo{author}{\bibfnamefont{C.}~\bibnamefont{Hauert}},
  \bibinfo{journal}{Phys. Rev. E} \textbf{\bibinfo{volume}{79}},
  \bibinfo{pages}{046707} (\bibinfo{year}{2009}).

\bibitem[{\citenamefont{Gr{\"{o}}nlund and Holme}(2005)}]{gh05}
\bibinfo{author}{\bibfnamefont{A.}~\bibnamefont{Gr{\"{o}}nlund}}
  \bibnamefont{and} \bibinfo{author}{\bibfnamefont{P.}~\bibnamefont{Holme}},
  \bibinfo{journal}{Adv. Complex Syst.} \textbf{\bibinfo{volume}{8}},
  \bibinfo{pages}{261} (\bibinfo{year}{2005}).

\bibitem[{\citenamefont{Pugliese and Castellano}(2009)}]{pc09}
\bibinfo{author}{\bibfnamefont{E.}~\bibnamefont{Pugliese}} \bibnamefont{and}
  \bibinfo{author}{\bibfnamefont{C.}~\bibnamefont{Castellano}},
  \bibinfo{journal}{Europhys. Lett.} \textbf{\bibinfo{volume}{88}},
  \bibinfo{pages}{58004} (\bibinfo{year}{2009}).

\bibitem[{\citenamefont{Szab\'{o} and F\'{a}th}(2007)}]{sf07}
\bibinfo{author}{\bibfnamefont{G.}~\bibnamefont{Szab\'{o}}} \bibnamefont{and}
  \bibinfo{author}{\bibfnamefont{G.}~\bibnamefont{F\'{a}th}},
  \bibinfo{journal}{Phys. Rep.} \textbf{\bibinfo{volume}{446}},
  \bibinfo{pages}{97} (\bibinfo{year}{2007}).

\bibitem[{\citenamefont{Guan et~al.}(2006)\citenamefont{Guan, Wu, Huang, Xu,
  and Wang}}]{gwhxw06}
\bibinfo{author}{\bibfnamefont{J.-Y.} \bibnamefont{Guan}},
  \bibinfo{author}{\bibfnamefont{Z.-X.} \bibnamefont{Wu}},
  \bibinfo{author}{\bibfnamefont{Z.-G.} \bibnamefont{Huang}},
  \bibinfo{author}{\bibfnamefont{X.-J.} \bibnamefont{Xu}}, \bibnamefont{and}
  \bibinfo{author}{\bibfnamefont{Y.-H.} \bibnamefont{Wang}},
  \bibinfo{journal}{Europhys. Lett.} \textbf{\bibinfo{volume}{76}},
  \bibinfo{pages}{1214} (\bibinfo{year}{2006}).

\bibitem[{\citenamefont{Kim et~al.}(2002)\citenamefont{Kim, Trusina, Holme,
  Minnhagen, Chung, and Choi}}]{kthmcc02}
\bibinfo{author}{\bibfnamefont{B.~J.} \bibnamefont{Kim}},
  \bibinfo{author}{\bibfnamefont{A.}~\bibnamefont{Trusina}},
  \bibinfo{author}{\bibfnamefont{P.}~\bibnamefont{Holme}},
  \bibinfo{author}{\bibfnamefont{P.}~\bibnamefont{Minnhagen}},
  \bibinfo{author}{\bibfnamefont{J.~S.} \bibnamefont{Chung}}, \bibnamefont{and}
  \bibinfo{author}{\bibfnamefont{M.~Y.} \bibnamefont{Choi}},
  \bibinfo{journal}{Phys. Rev. E} \textbf{\bibinfo{volume}{66}},
  \bibinfo{pages}{021907} (\bibinfo{year}{2002}).

\bibitem[{\citenamefont{Ohtsuki
  et~al.}(2007{\natexlab{a}})\citenamefont{Ohtsuki, Nowak, and
  Pacheco}}]{onp07}
\bibinfo{author}{\bibfnamefont{H.}~\bibnamefont{Ohtsuki}},
  \bibinfo{author}{\bibfnamefont{M.~A.} \bibnamefont{Nowak}}, \bibnamefont{and}
  \bibinfo{author}{\bibfnamefont{J.~M.} \bibnamefont{Pacheco}},
  \bibinfo{journal}{Phys. Rev. Lett.} \textbf{\bibinfo{volume}{98}},
  \bibinfo{pages}{108106} (\bibinfo{year}{2007}{\natexlab{a}}).

\bibitem[{\citenamefont{Ohtsuki
  et~al.}(2007{\natexlab{b}})\citenamefont{Ohtsuki, Pacheco, and
  Nowak}}]{opn07}
\bibinfo{author}{\bibfnamefont{H.}~\bibnamefont{Ohtsuki}},
  \bibinfo{author}{\bibfnamefont{J.~M.} \bibnamefont{Pacheco}},
  \bibnamefont{and} \bibinfo{author}{\bibfnamefont{M.~A.} \bibnamefont{Nowak}},
  \bibinfo{journal}{J. Theor. Biol.} \textbf{\bibinfo{volume}{246}},
  \bibinfo{pages}{681} (\bibinfo{year}{2007}{\natexlab{b}}).

\bibitem[{\citenamefont{Szolnoki and Szab\'{o}}(2007)}]{ss07}
\bibinfo{author}{\bibfnamefont{A.}~\bibnamefont{Szolnoki}} \bibnamefont{and}
  \bibinfo{author}{\bibfnamefont{G.}~\bibnamefont{Szab\'{o}}},
  \bibinfo{journal}{Europhys. Lett.} \textbf{\bibinfo{volume}{77}},
  \bibinfo{pages}{30004} (\bibinfo{year}{2007}).

\bibitem[{\citenamefont{Wu et~al.}(2006)\citenamefont{Wu, Xu, and
  Wang}}]{wxw06}
\bibinfo{author}{\bibfnamefont{Z.-X.} \bibnamefont{Wu}},
  \bibinfo{author}{\bibfnamefont{X.-J.} \bibnamefont{Xu}}, \bibnamefont{and}
  \bibinfo{author}{\bibfnamefont{Y.-H.} \bibnamefont{Wang}},
  \bibinfo{journal}{Chin. Phys. Lett.} \textbf{\bibinfo{volume}{23}},
  \bibinfo{pages}{531} (\bibinfo{year}{2006}).

\bibitem[{\citenamefont{Boccaletti et~al.}(2006)\citenamefont{Boccaletti,
  Latora, Moreno, Chavez, and Hwang}}]{blmch06}
\bibinfo{author}{\bibfnamefont{S.}~\bibnamefont{Boccaletti}},
  \bibinfo{author}{\bibfnamefont{V.}~\bibnamefont{Latora}},
  \bibinfo{author}{\bibfnamefont{Y.}~\bibnamefont{Moreno}},
  \bibinfo{author}{\bibfnamefont{M.}~\bibnamefont{Chavez}}, \bibnamefont{and}
  \bibinfo{author}{\bibfnamefont{D.-U.} \bibnamefont{Hwang}},
  \bibinfo{journal}{Phys. Rep.} \textbf{\bibinfo{volume}{424}},
  \bibinfo{pages}{175} (\bibinfo{year}{2006}).

\bibitem[{\citenamefont{Bornholdt and Schuster}(2003)}]{bs03}
\bibinfo{editor}{\bibfnamefont{S.}~\bibnamefont{Bornholdt}} \bibnamefont{and}
  \bibinfo{editor}{\bibfnamefont{H.~G.} \bibnamefont{Schuster}}, eds.,
  \emph{\bibinfo{title}{Handbook of Graphs and Networks}}
  (\bibinfo{publisher}{Wiley-VCH}, \bibinfo{address}{Weinheim, Germany},
  \bibinfo{year}{2003}).

\bibitem[{\citenamefont{Newman et~al.}(2006)\citenamefont{Newman, Barab\'{a}si,
  and Watts}}]{nbw06}
\bibinfo{editor}{\bibfnamefont{M.}~\bibnamefont{Newman}},
  \bibinfo{editor}{\bibfnamefont{A.-L.} \bibnamefont{Barab\'{a}si}},
  \bibnamefont{and} \bibinfo{editor}{\bibfnamefont{D.~J.} \bibnamefont{Watts}},
  eds., \emph{\bibinfo{title}{The Structure and Dynamics of Networks}}
  (\bibinfo{publisher}{Princeton University Press},
  \bibinfo{address}{Princeton, NJ}, \bibinfo{year}{2006}).

\bibitem[{\citenamefont{Antal et~al.}(2006)\citenamefont{Antal, Redner, and
  Sood}}]{ars06}
\bibinfo{author}{\bibfnamefont{T.}~\bibnamefont{Antal}},
  \bibinfo{author}{\bibfnamefont{S.}~\bibnamefont{Redner}}, \bibnamefont{and}
  \bibinfo{author}{\bibfnamefont{V.}~\bibnamefont{Sood}},
  \bibinfo{journal}{Phys. Rev. Lett.} \textbf{\bibinfo{volume}{96}},
  \bibinfo{pages}{188104} (\bibinfo{year}{2006}).

\bibitem[{\citenamefont{Broom and Rycht\'{a}\v{r}}(2008)}]{br08}
\bibinfo{author}{\bibfnamefont{M.}~\bibnamefont{Broom}} \bibnamefont{and}
  \bibinfo{author}{\bibfnamefont{J.}~\bibnamefont{Rycht\'{a}\v{r}}},
  \bibinfo{journal}{P. Roy. Soc. A-Math. Phy.} \textbf{\bibinfo{volume}{464}},
  \bibinfo{pages}{2609} (\bibinfo{year}{2008}).

\bibitem[{\citenamefont{Masuda and Ohtsuki}(2009)}]{mo09}
\bibinfo{author}{\bibfnamefont{N.}~\bibnamefont{Masuda}} \bibnamefont{and}
  \bibinfo{author}{\bibfnamefont{H.}~\bibnamefont{Ohtsuki}},
  \bibinfo{journal}{New J. Phys.} \textbf{\bibinfo{volume}{11}},
  \bibinfo{pages}{033012} (\bibinfo{year}{2009}).

\bibitem[{\citenamefont{Sood et~al.}(2008)\citenamefont{Sood, Antal, and
  Redner}}]{sar08}
\bibinfo{author}{\bibfnamefont{V.}~\bibnamefont{Sood}},
  \bibinfo{author}{\bibfnamefont{T.}~\bibnamefont{Antal}}, \bibnamefont{and}
  \bibinfo{author}{\bibfnamefont{S.}~\bibnamefont{Redner}},
  \bibinfo{journal}{Phys. Rev. E} \textbf{\bibinfo{volume}{77}},
  \bibinfo{pages}{041121} (\bibinfo{year}{2008}).

\bibitem[{\citenamefont{Barbosa et~al.}(2009)\citenamefont{Barbosa, Donangelo,
  and Souza}}]{bds09}
\bibinfo{author}{\bibfnamefont{V.~C.} \bibnamefont{Barbosa}},
  \bibinfo{author}{\bibfnamefont{R.}~\bibnamefont{Donangelo}},
  \bibnamefont{and} \bibinfo{author}{\bibfnamefont{S.~R.} \bibnamefont{Souza}},
  \bibinfo{journal}{Phys. Rev. E} \textbf{\bibinfo{volume}{80}},
  \bibinfo{pages}{026115} (\bibinfo{year}{2009}).

\bibitem[{\citenamefont{Erd\H{o}s and R\'{e}nyi}(1959)}]{er59}
\bibinfo{author}{\bibfnamefont{P.}~\bibnamefont{Erd\H{o}s}} \bibnamefont{and}
  \bibinfo{author}{\bibfnamefont{A.}~\bibnamefont{R\'{e}nyi}},
  \bibinfo{journal}{Publ. Math. (Debrecen)} \textbf{\bibinfo{volume}{6}},
  \bibinfo{pages}{290} (\bibinfo{year}{1959}).

\bibitem[{\citenamefont{Karp}(1990)}]{k90}
\bibinfo{author}{\bibfnamefont{R.~M.} \bibnamefont{Karp}},
  \bibinfo{journal}{Random Struct. Algor.} \textbf{\bibinfo{volume}{1}},
  \bibinfo{pages}{73} (\bibinfo{year}{1990}).

\bibitem[{\citenamefont{Barbosa et~al.}(2003)\citenamefont{Barbosa, Donangelo,
  and Souza}}]{bds03}
\bibinfo{author}{\bibfnamefont{V.~C.} \bibnamefont{Barbosa}},
  \bibinfo{author}{\bibfnamefont{R.}~\bibnamefont{Donangelo}},
  \bibnamefont{and} \bibinfo{author}{\bibfnamefont{S.~R.} \bibnamefont{Souza}},
  \bibinfo{journal}{Physica A} \textbf{\bibinfo{volume}{321}},
  \bibinfo{pages}{381} (\bibinfo{year}{2003}).

\bibitem[{\citenamefont{{El-Shehawey}}(2000)}]{e00}
\bibinfo{author}{\bibfnamefont{M.~A.} \bibnamefont{{El-Shehawey}}},
  \bibinfo{journal}{J. Phys. A-Math. Gen.} \textbf{\bibinfo{volume}{33}},
  \bibinfo{pages}{9005} (\bibinfo{year}{2000}).

\bibitem[{\citenamefont{Nowak}(2006)}]{n06}
\bibinfo{author}{\bibfnamefont{M.~A.} \bibnamefont{Nowak}},
  \emph{\bibinfo{title}{Evolutionary Dynamics}} (\bibinfo{publisher}{Harvard
  University Press}, \bibinfo{address}{Cambridge, MA}, \bibinfo{year}{2006}).

\end{thebibliography}
\bibliographystyle{apsrev}

\end{document}